\documentclass{llncs}
\usepackage{amsfonts}
\usepackage{feynmf}             %Package for feynman diagrams.
\begin{document}

\title{Can a computer be ``pushed'' to perform faster-than-light?}
\titlerunning{Can a computer be ``pushed'' to perform faster-than-light?}

%\author{Volkmar Putz}
%\author{Karl Svozil}
%\email{svozil@tuwien.ac.at}
%\homepage{http://tph.tuwien.ac.at/~svozil}
%\affiliation{Institute for Theoretical Physics, Vienna University of Technology,  \\ Wiedner Hauptstra\ss e 8-10/136, A-1040 Vienna, Austria}

\author{Volkmar Putz and Karl Svozil}
\institute{Institute for Theoretical Physics, Vienna University of Technology,  \\
Wiedner Hauptstra\ss e 8-10/136, A-1040 Vienna, Austria \\
{http://tph.tuwien.ac.at/\homedir svozil}\\
\email{putz@hep.itp.tuwien.ac.at,svozil@tuwien.ac.at}
}

%\pacs{03.67.Hk,03.65.Ud}
%\keywords{Quantum information, quantum recursion theory, Thomson lamp, hypercomputation, accelerated computers}

\maketitle

\begin{abstract}
We propose to ``boost'' the speed of communication and computation by immersing the computing environment into a medium whose index of refraction is smaller than one, thereby trespassing the speed-of-light barrier.
\end{abstract}

The Church-Turing and the Cook-Karp theses, as well as other,
more general limits on computation, are under permanent ``scrutiny'' (cf., e.g., Ref.~\cite[p~11]{davis-58} or Ref.~\cite[p.~5]{deutsch}) by the physical sciences.
Some recent issues which have been raised comprise Zeno-squeezed accelerated time scales~\cite{weyl:49,Hogarth92,DBLP:conf/mcu/Durand-Lose04,Nemeti2006118,1612095}
enabling the construction of ``infinity machines'' capable of hypercomputation~\cite{Davis-2006,Doria-2006,ord-2006},
counterfactual computation~\cite{GRAEMEMITCHISON05082001} and cryptography~\cite{PhysRevLett.103.230501} based on quantum counterfactuals~\cite{elitzur-vaidman:1,vaidman:2009},
as well as the dissipation limits to computation~\cite{maxwell-demon}.
Here we shall consider the possibility to speed up optical~\cite{Chiao:02} computations and communication by transgressing the speed of light barrier in vacuum.
Note that, although the speed of light barrier appears to be a fundamental limit for the transfer of ``freely willable'' information~\cite{recami:01},
several ways for ``signals'' trespassing the relativistic light cone~\cite{0264-9381-11-5-001},
even to the extent of time travel~\cite{godel-sch,nahin,PhysRevD.46.603,deutsch91,svozil-greenberger-2005}, have been proposed.
There appears to be a consensus that, just as for quantum correlations featuring (un)controllable non-locality~\cite{shimony2}
{\it via} outcome dependence but parameter independence, ``signal'' signatures beyond the velocity of light limit~\cite{0953-4075-35-6-201}
could be tolerated at the kinematical level~\cite{Liberati2002167} as long as they are ``benign''
and thus incapable of rendering diagonalization-type~\cite{davis-58,smullyan-92} paradoxes.
This means that no paradoxes of self-referentiality, such as the ``grandfather paradox'' (e.g.,
by  travelling back in time and killing one's own biological grandfather before the latter has met one's grandmother),
should occur~\cite{bell-j-l-paradox}.

In what follows we propose to ``boost'' the speed of communication and computation
by ``pushing'' the computer into a medium whose index of refraction is smaller than one.
The speed of communication by light signals varies indirectly proportional to the index of refraction,
differing  greatly for various forms of media,
substrata or ``ethers'' susceptible of the traversal of light.
Quantum field theory allows the index of refraction to become smaller than one, thereby
formally indicating a speed of photons exceeding the classical speed of light limit in vacuum.

How can one envisage such a computational substratum?
One concrete realization would be the construction of an universal optical computer based on
beam splitters~\cite{zeilinger:882}
capable of rendering arbitrary discrete unitary transformations~\cite{rzbb,zukowski-97,svozil-2004-analog} immersed in a transparent medium occupied by charged fermions.
Note that, as optical computers are far more than just photons or beams of light,
a necessary requirement for any such computer to properly function
would be that the optical components of the computer, such as in particular beam splitters and phase shifters,
would work as expected in such a medium.

``Diagrammatically speaking''~\cite{feynman-62,schweber-62,thooft-Veltman_Diagrammar}, i.e., in terms of perturbative quantum field theory,
a photon, i.e., the ``unit quantum of light''
associated with a particular mode of the electromagnetic field,
travels through the vacuum ether medium~\cite{dirac-aether} by polarizing it
through partly ``splitting up'' into an electron-positron pair and recombining.
In solid state physics, this phenomenon gives rise to lattice excitations called {\em phonons}~\cite{stroscio}.
The electrons and positrons are themselves subject to higher order radiative corrections involving photons.

Thus, any change of vacuum polarization, such as finite boundary conditions, or increased or decreased pair production,
alters the susceptibility of the vacuum ether medium for carrying electromagnetic waves,
and thus results in a change of the velocity of light.
Historically, this effect has first been studied
for magnetic fields~\cite{er:61,RevModPhys.38.626,Adler1971599} and finite temperatures~\cite{Gies1998420}.
The first indication of a vacuum polarization-induced index of refraction {\em smaller than one} was reported by
Scharnhorst~\cite{scharnhorst,milonni,Scharnhorst-1998} and Barton~\cite{barton,0305-4470-26-8-024}
in an attempt to utilize the reduced vacuum polarization
in the ``Casimir vacuum''~\cite{milonni-book} between two conducting parallel plates.
More recently, trans-vacuum-speed
metamaterials~\cite{PhysRevE.63.046604,PhysRevE.68.026612,PhysRevE.70.068601,PhysRevE.70.068602,PhysRevE.78.016601}
as well as negative refractive indices in gyrotropically magnetoelectric media~\cite{PhysRevB.75.196101} have been suggested.
It would be interesting to extend these calculations to the squeezed vacuum state by computing the polarization in such an ``exotic'' vacuum~\cite{svozil-putz}.

One of the possibilities which have not been discussed so far is the immersion of the computing environment into a vacuum ether medium ``filled'' with electrons or positrons.
In such an environment, the Pauli exclusion principle would ``attenuate'' pair creation, thereby reducing the  polarization of the medium, resulting in
a reduced index of refraction as well as in an increase of the velocity of light.

\begin{figure}
\begin{center}
\begin{fmffile}{QED_vacuum_polarization}
\begin{fmfgraph*}(120,40)
\fmfleft{i}
\fmfright{o}
\fmf{photon,label=$k$}{i,v1}
\fmf{photon,label=$k$}{v2,o}
\fmf{fermion,left,tension=0.4}{v1,v2,v1}
%\fmf{photon}{v1,v2}
\fmfdot{v1,v2}
\put(30.00,-10.00){\framebox(60,60)[cc]{$\varepsilon_F$}}
\end{fmfgraph*}
\end{fmffile}
\end{center}
\caption{Lowest order vacuum polarization diagram.\label{2010-uc10-f1}}
\end{figure}
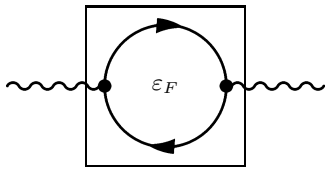

After regularization and renormalization, the lowest order  change
to the radiative correction associated with  the vacuum polarization  (whose Feynman diagram is depicted in Fig.~\ref{2010-uc10-f1})
can be written as~\cite{RevModPhys.21.434,PhysRev.76.769,schweber-62}
\begin{equation}
{\rm \Delta }{\bf \Pi}_{\mu \nu}(k^2)=-\left(g_{\mu \nu }k^2 - k_\mu k_\nu \right) \frac{2\alpha}{3\pi}  \log \frac{\varepsilon_F}{m},
\end{equation}
where $m$ stands for the electron rest mass and $\varepsilon_F$ denotes the cutoff
associated with the filled electron or positron modes; the calculation assumed $k^2<m$.
Let $\epsilon_\mu$ stand for the vacuum polarization.
Then we can introduce an effective mass term~\cite{PhysRev.82.664,PhysRevD.10.492,PhysRevD.12.1132}
\begin{equation}
M(k)=\epsilon^\mu {\bf \Pi}_{\mu \nu}(k)\epsilon^\nu
\end{equation}
such that the eigenvalue equation is
\begin{equation}
{\bf k}^2+ M(k)=(k^0)^2,
\end{equation}
where $k^\mu=({\bf k},k^0=\omega)$; and
\begin{equation}
\vert  {\bf k} \vert \approx \omega - \frac{1}{2 \omega} M(k).
\end{equation}
Thus the index of refraction can be defined by
\begin{equation}
n(\omega )=\frac{\vert {\bf k} \vert}{\omega}\approx 1 - \frac{1}{2 \omega^2} M(k).
\end{equation}
Hence the change of the refractive index is given by
\begin{equation}
{\rm \Delta }n(\omega )\approx -\frac{\alpha}{3\pi \omega^2} (\epsilon^\mu k_\mu)^2  \log \frac{\varepsilon_F}{m}.
\label{2010-uc10-e1}
\end{equation}

The group velocity is given by~\cite[Equ.~(2)]{Scharnhorst-1998} $v_{gr}=c/n_{gr}$ with
$n_{gr}(\omega )= n (\omega )+ \omega \left[\partial n (\omega )/\partial \omega \right]$,
which, for transversal waves, turns out to be $n (\omega )$.
As a result, the speed of light $c/(1-{\rm \Delta }n)\approx c+{\rm \Delta }c$  is changed by  ${\rm \Delta }c = c {\rm \Delta }n$.

Note that
group velocities,
like phase velocities and energy velocities, are not in general signal velocities.
Thus a group velocity exceeding the vacuum speed of light $c$ does
not contradict relativity
\cite{PhysRevA.48.R34,Diener1996327,Chiao:02}.

Nevertheless, as has already pointed out, this effect can be used to ``push'' the computer into a domain of faster-than-light
computation; with the possibility to decrease its time cycles accordingly.
One should keep in mind that at present such a possibility merely remains a theoretical speculation;
this hypothetical character being shared with some relativistic ``realizations'' of hypercomputers.
Nevertheless it might be interesting to pursue the possibilities related to temporal quantum field theoretical speedup further,
for in principle nothing prevents ${\rm \Delta }n$ in Eq.~(\ref{2010-uc10-e1}) or in other ``exotic'' vacuum states from approaching one,
yielding an unbounded cycle speed, associated with expanding memory requirements~\cite{calude-staiger-09}.

In summary we have discussed field theoretic options for the ``speedup'' of communication and computation.
These are based on the alteration of the polarization of ``exotic vacua'' and the respective changes of the index of refraction.
The  speed of light is modified in indirect proportion to the refractive index of the medium it is travelling through.
Thus for materials with a refractive index smaller than unity,
light travels faster than it does in  ``normal''  vacuum whose index of refraction is associated with unity.
Hence, optical computers operating in such an ``exotic''  medium, if they existed, could compute faster
than computers in ``normal'' vacuum or ordinary materials which have refractive indices equal to or greater than unity.
Feasible realization of universal computers utilizing this effect
could employ generalized beam splitters capable of realizing arbitrary discrete unitary operators.

We have discussed a general  physical framework for ``exotic'' vacua with indices of refraction strictly smaller than unity.
One such vacuum state is responsible for the hypothetical
Scharnhorst effect, for which the polarizability of the vacuum ``medium''
is effectively reduced by the boundary conditions of the electromagnetic field between two conductors (e.g., parallel plates).
Another possibility which is introduced here is the occupancy of charged fermionic, in particular electronic, states,
which would partially inhibit the pair production of fermion-antifermion  (electron-positron) pairs contributing to the
vacuum polarization even in lowest nontrivial order of the perturbation series.
It should be emphasized that these findings do not represent the possibility to circumvent relativistic causality,
nor are they inconsistent with the present formalism of relativity theory or the theory of quantized fields.

%\bibliography{svozil}

\begin{thebibliography}{10}

\bibitem{davis-58}
Davis, M.:
\newblock Computability and Unsolvability.
\newblock McGraw-Hill, New York (1958)

\bibitem{deutsch}
Deutsch, D.:
\newblock Quantum theory, the {C}hurch-{T}uring principle and the universal
  quantum computer.
\newblock Proceedings of the Royal Society of London. Series A, Mathematical
  and Physical Sciences (1934-1990) \textbf{400} (1985)  97--117

\bibitem{weyl:49}
Weyl, H.:
\newblock Philosophy of Mathematics and Natural Science.
\newblock Princeton University Press, Princeton (1949)

\bibitem{Hogarth92}
Hogarth, M.L.:
\newblock Does general relativity allow an observer to view an eternity in a
  finite time?
\newblock Foundations of Physics Letters \textbf{5} (1992)  173--181

\bibitem{DBLP:conf/mcu/Durand-Lose04}
Durand-Lose, J.:
\newblock Abstract geometrical computation for black hole computation.
\newblock In Margenstern, M., ed.: Machines, Computations, and Universality,
  4th International Conference, MCU 2004, Saint Petersburg, Russia, September
  21-24, 2004, Revised Selected Papers. Volume 3354 of Lecture Notes in
  Computer Science., Springer (2005)  176--187

\bibitem{Nemeti2006118}
N{\'{e}}meti, I., D{\'{a}}vid, G.:
\newblock Relativistic computers and the {T}uring barrier.
\newblock Applied Mathematics and Computation \textbf{178} (2006)  118--142
  Special Issue on Hypercomputation.

\bibitem{1612095}
Svozil, K.:
\newblock On the brightness of the {T}homson lamp: A prolegomenon to quantum
  recursion theory.
\newblock In Calude, C.S., Costa, J.F., Dershowitz, N., Freire, E., Rozenberg,
  G., eds.: UC '09: Proceedings of the 8th International Conference on
  Unconventional Computation, Berlin, Heidelberg, Springer Verlag (2009)
  236--246

\bibitem{Davis-2006}
Davis, M.:
\newblock Why there is no such discipline as hypercomputation.
\newblock Applied Mathematics and Computation \textbf{178} (2006)  4--7

\bibitem{Doria-2006}
Doria, F.A., Costa, J.F.:
\newblock Introduction to the special issue on hypercomputation.
\newblock Applied Mathematics and Computation \textbf{178} (2006)  1--3

\bibitem{ord-2006}
Ord, T.:
\newblock The many forms of hypercomputation.
\newblock Applied Mathematics and Computation \textbf{178} (2006)  143--153

\bibitem{GRAEMEMITCHISON05082001}
Mitchison, G., Jozsa, R.:
\newblock Counterfactual computation.
\newblock Proceedings of the Royal Society of London. Series A: Mathematical,
  Physical and Engineering Sciences \textbf{457} (2001)  1175--1193

\bibitem{PhysRevLett.103.230501}
Noh, T.G.:
\newblock Counterfactual quantum cryptography.
\newblock Physical Review Letters \textbf{103} (2009)  230501

\bibitem{elitzur-vaidman:1}
Elitzur, A.C., Vaidman, L.:
\newblock Quantum mechanical interaction-free measurements.
\newblock Foundations of Physics \textbf{23} (1993)  987--997

\bibitem{vaidman:2009}
Vaidman, L.:
\newblock Counterfactuals in quantum mechanics.
\newblock In Greenberger, D., Hentschel, K., Weinert, F., eds.: Compendium of
  Quantum Physics.
\newblock Springer, Berlin, Heidelberg (2007)  132--136

\bibitem{maxwell-demon}
Leff, H.S., Rex, A.F.:
\newblock Maxwell's Demon.
\newblock Princeton University Press, Princeton (1990)

\bibitem{Chiao:02}
Chiao, R.Y., Milonni, P.W.:
\newblock Fast light, slow light.
\newblock Optics \& Photonics News \textbf{13} (2002)  26--30

\bibitem{recami:01}
Recami, E.:
\newblock Superluminal motions? {A} bird's--eye view of the experimental
  situation.
\newblock Foundation of Physics \textbf{31} (2001)  1119--1135

\bibitem{0264-9381-11-5-001}
Alcubierre, M.:
\newblock The warp drive: hyper-fast travel within general relativity.
\newblock Classical and Quantum Gravity \textbf{11} (1994)  L73--L77

\bibitem{godel-sch}
G{\"{o}}del, K.:
\newblock A remark about the relationship between relativity theory and
  idealistic philosophy.
\newblock In Schilpp, P.A., ed.: {A}lbert {E}instein,
  {P}hilosopher-{S}cientist.
\newblock Tudor Publishing Company, New York (1949)  555--561 Reprinted in
  Ref.~\cite[pp. 202-207]{godel-ges2}.

\bibitem{nahin}
Nahin, P.J.:
\newblock Time Travel (Second edition).
\newblock AIP Press and Springer, New York (1998)

\bibitem{PhysRevD.46.603}
W.Hawking, S.:
\newblock Chronology protection conjecture.
\newblock Physical Review D \textbf{46} (1992)  603--611

\bibitem{deutsch91}
Deutsch, D.:
\newblock Quantum mechanics near closed timelike lines.
\newblock Physical Review D \textbf{44} (1991)  3197--3217

\bibitem{svozil-greenberger-2005}
Greenberger, D.M., Svozil, K.:
\newblock Quantum theory looks at time travel.
\newblock In A.~Elitzur, S.D., Kolenda, N., eds.: Quo Vadis Quantum Mechanics?,
  Berlin, Springer Verlag (2005)  63--72

\bibitem{shimony2}
Shimony, A.:
\newblock Controllable and uncontrollable non-locality.
\newblock In {\it et al.}, S.K., ed.: Proceedings of the International
  Symposium on the Foundations of Quantum Mechanics, Tokyo, Physical Society of
  Japan (1984)  225--230 See also J. Jarrett, {\sl Bell's Theorem, Quantum
  Mechanics and Local Realism}, Ph. D. thesis, Univ. of Chicago, 1983; {\sl
  Nous}, {\bf 18}, 569 (1984).

\bibitem{0953-4075-35-6-201}
Milonni, P.W.:
\newblock Controlling the speed of light pulses.
\newblock Journal of Physics B: Atomic, Molecular and Optical Physics
  \textbf{35} (2002)  R31--R56

\bibitem{Liberati2002167}
Liberati, S., Sonego, S., Visser, M.:
\newblock Faster-than-c signals, special relativity, and causality.
\newblock Annals of Physics \textbf{298} (2002)  167--185

\bibitem{smullyan-92}
Smullyan, R.M.:
\newblock {G}{\"{o}}del's Incompleteness Theorems.
\newblock Oxford University Press, New York, New York (1992)

\bibitem{bell-j-l-paradox}
Bell, J.L.:
\newblock Time and causation in {G}{\"o}del's universe.
\newblock Transcendent Philosophy \textbf{3} (2002) ~1

\bibitem{zeilinger:882}
Zeilinger, A.:
\newblock General properties of lossless beam splitters in interferometry.
\newblock American Journal of Physics \textbf{49} (1981)  882--883

\bibitem{rzbb}
Reck, M., Zeilinger, A., Bernstein, H.J., Bertani, P.:
\newblock Experimental realization of any discrete unitary operator.
\newblock Physical Review Letters \textbf{73} (1994)  58--61

\bibitem{zukowski-97}
Zukowski, M., Zeilinger, A., Horne, M.A.:
\newblock Realizable higher-dimensional two-particle entanglements via
  multiport beam splitters.
\newblock Physical Review A (Atomic, Molecular, and Optical Physics)
  \textbf{55} (1997)  2564--2579

\bibitem{svozil-2004-analog}
Svozil, K.:
\newblock Noncontextuality in multipartite entanglement.
\newblock J. Phys. A: Math. Gen. \textbf{38} (2005)  5781--5798

\bibitem{feynman-62}
Feynman, R.P.:
\newblock Quantum Electrodynamics.
\newblock Addison-Wesley, Redwood City, CA (1962)

\bibitem{schweber-62}
Schweber, S.:
\newblock Relativistic Quantum Field Theory.
\newblock Harper and Row, New York (1984)

\bibitem{thooft-Veltman_Diagrammar}
`t~Hooft, G., Veltman, M.:
\newblock Diagrammar.
\newblock CERN preprint 73-9 (1973)

\bibitem{dirac-aether}
Dirac, P.A.M.:
\newblock Is there an aether?
\newblock Nature \textbf{168} (1951)  906--907

\bibitem{stroscio}
Stroscio, M.A., Dutta, M.:
\newblock Phonons in Nanostructures.
\newblock Cambridge University Press, Cambridge (2005)

\bibitem{er:61}
Erber, T.:
\newblock Velocity of light in a magnetic field.
\newblock Nature \textbf{190} (1961)  25--27

\bibitem{RevModPhys.38.626}
Erber, T.:
\newblock High-energy electromagnetic conversion processes in intense magnetic
  fields.
\newblock Reviews of Modern Physics \textbf{38} (1966)  626--659

\bibitem{Adler1971599}
Adler, S.L.:
\newblock Photon splitting and photon dispersion in a strong magnetic field.
\newblock Annals of Physics \textbf{67} (1971)  599--647

\bibitem{Gies1998420}
Gies, H., Dittrich, W.:
\newblock Light propagation in non-trivial {QED} vacua.
\newblock Physics Letters B \textbf{431} (1998)  420 -- 429

\bibitem{scharnhorst}
Scharnhorst, K.:
\newblock On propagation of light in the vacuum between plates.
\newblock Physics Letters B \textbf{236} (1990)  354--359

\bibitem{milonni}
Milonni, P., Svozil, K.:
\newblock Impossibility of measuring faster-than-c signaling by the
  {S}charnhorst effect.
\newblock Physics Letters B \textbf{248} (1990)  437--438

\bibitem{Scharnhorst-1998}
Scharnhorst, K.:
\newblock The velocities of light in modified {QED} vacua.
\newblock Annalen der Physik \textbf{7} (1998)  700--709

\bibitem{barton}
Barton, G.:
\newblock Faster-than-c light between parallel mirrors. {T}he {S}charnhorst
  effect rederived.
\newblock Physics Letters B \textbf{237} (1990)  559--562

\bibitem{0305-4470-26-8-024}
Barton, G., Scharnhorst, K.:
\newblock {QED} between parallel mirrors: light signals faster than c, or
  amplified by the vacuum.
\newblock Journal of Physics A: Mathematical and General \textbf{26} (1993)
  2037--2046

\bibitem{milonni-book}
Milonni, P.W.:
\newblock The Quantum Vacuum.
\newblock Academic Press, San Diego (1994)

\bibitem{PhysRevE.63.046604}
Ziolkowski, R.W.:
\newblock Superluminal transmission of information through an electromagnetic
  metamaterial.
\newblock Physical Review E \textbf{63} (2001)  046604

\bibitem{PhysRevE.68.026612}
Ziolkowski, R.W., Cheng, C.Y.:
\newblock Existence and design of trans-vacuum-speed metamaterials.
\newblock Physical Review E \textbf{68} (2003)  026612

\bibitem{PhysRevE.70.068601}
Tretyakov, S.A.:
\newblock Comment on ``existence and design of trans-vacuum-speed
  metamaterials''.
\newblock Physical Review E \textbf{70} (2004)  068601

\bibitem{PhysRevE.70.068602}
Ziolkowski, R.W.:
\newblock Reply to ``comment on `existence and design of trans-vacuum-speed
  metamaterials' ''.
\newblock Physical Review E \textbf{70} (2004)  068602

\bibitem{PhysRevE.78.016601}
Shvartsburg, A.B., Marklund, M., Brodin, G., Stenflo, L.:
\newblock Superluminal tunneling of microwaves in smoothly varying transmission
  lines.
\newblock Physical Review E \textbf{78} (2008)  016601

\bibitem{PhysRevB.75.196101}
Qiu, C.W., Zouhdi, S.:
\newblock Comment on ``negative refractive index in gyrotropically
  magnetoelectric media''.
\newblock Phys. Rev. B \textbf{75} (2007)  196101

\bibitem{svozil-putz}
Putz, V., Svozil, K.:
\newblock Quantum electrodynamics in the squeezed vacuum state: electron mass
  shift.
\newblock Il Nuovo Cimento B \textbf{119} (2004)  175--179

\bibitem{RevModPhys.21.434}
Pauli, W., Villars, F.:
\newblock On the invariant regularization in relativistic quantum theory.
\newblock Reviews of Modern Physics \textbf{21} (1949)  434--444

\bibitem{PhysRev.76.769}
Feynman, R.P.:
\newblock Space-time approach to quantum electrodynamics.
\newblock Physical Review \textbf{76} (1949)  769--789

\bibitem{PhysRev.82.664}
Schwinger, J.:
\newblock On gauge invariance and vacuum polarization.
\newblock Physical Review \textbf{82} (1951)  664--679

\bibitem{PhysRevD.10.492}
Tsai, W., Erber, T.:
\newblock Photon pair creation in intense magnetic fields.
\newblock Physical Review D \textbf{10} (1974)  492--499

\bibitem{PhysRevD.12.1132}
Tsai, W., Erber, T.:
\newblock Propagation of photons in homogeneous magnetic fields: Index of
  refraction.
\newblock Physical Review D \textbf{12} (1975)  1132--1137

\bibitem{PhysRevA.48.R34}
Chiao, R.Y.:
\newblock Superluminal (but causal) propagation of wave packets in transparent
  media with inverted atomic populations.
\newblock Phys. Rev. A \textbf{48} (1993)  R34--R37

\bibitem{Diener1996327}
Diener, G.:
\newblock Superluminal group velocities and information transfer.
\newblock Physics Letters A \textbf{223} (1996)  327 -- 331

\bibitem{calude-staiger-09}
Calude, C.S., Staiger, L.:
\newblock A note on accelerated {T}uring machines.
\newblock {CDMTCS} preprint nr.~350, 7~p. (2009)

\bibitem{godel-ges2}
G{\"{o}}del, K.
\newblock In Feferman, S., {Dawson, Jr.}, J.W., Kleene, S.C., Moore, G.H.,
  Solovay, R.M., van Heijenoort, J., eds.: Collected Works. Publications
  1938-1974. Volume {II}.
\newblock Oxford University Press, Oxford (1990)

\end{thebibliography}

\end{document}